\documentstyle[12pt]{article}
\renewcommand{\thefootnote}{\fnsymbol{footnote}}

\newlength{\extraspace}
\newlength{\extraspaces}
\newcommand{\be}{\begin{equation}
\addtolength{\abovedisplayskip}{\extraspaces}
\addtolength{\belowdisplayskip}{\extraspaces}
\addtolength{\abovedisplayshortskip}{\extraspace}
\addtolength{\belowdisplayshortskip}{\extraspace}}
\newcommand{\ee}{\end{equation}}
\newcommand{\ba}{\begin{eqnarray}
\addtolength{\abovedisplayskip}{\extraspaces}
\addtolength{\belowdisplayskip}{\extraspaces}
\addtolength{\abovedisplayshortskip}{\extraspace}
\addtolength{\belowdisplayshortskip}{\extraspace}}
\newcommand{\ea}{\end{eqnarray}}
\newcommand{\newsection}[1]{
\vspace{7mm}
\pagebreak[3]
\addtocounter{section}{1}
\setcounter{equation}{0}
\setcounter{subsection}{0}

{\large {\bf \thesection. #1}}
\nopagebreak
\medskip
\nopagebreak
\hspace{3mm}}
\newcommand{\nonu}{\nonumber \\[.5mm]}
\newcommand{\A}{&\!\!\!}

\begin{document}
\addtolength{\baselineskip}{3.0mm}
\begin{flushright}
SIT-HEP/MT-1 \\
STUPP-00-160 \\ March, 2000
\end{flushright}
%

\begin{center}
{\large{\bf{Canonical formulation of 
            $N = 2$ supergravity 
            
            in terms of the Ashtekar variable}}} 
\\[2mm]
{\large Motomu Tsuda}
\footnote{e-mail: tsuda@sit.ac.jp} 
\\
Laboratory of Physics, Saitama Institute of Technology \\
Okabe-machi, Saitama 369-0293, Japan 

{\large and}

{\large Takeshi Shirafuji}
\footnote{e-mail: sirafuji@post.saitama-u.ac.jp} 
\\
Physics Department, Saitama University \\
Urawa, Saitama 338-8570, Japan \\[1mm]
{\bf Abstract}\\[1mm]
{\parbox{13cm}{\hspace{5mm} 
We reconstruct the Ashtekar's canonical 
formulation of $N = 2$ supergravity (SUGRA) starting 
from the $N = 2$ chiral Lagrangian derived by closely 
following the method employed in the usual SUGRA. 
In order to get the full graded algebra of the Gauss, 
$U(1)$ gauge and right-handed supersymmetry (SUSY) 
constraints, we extend the internal, global $O(2)$ 
invariance to local one by introducing a cosmological 
constant to the chiral Lagrangian. 
The resultant Lagrangian does not contain any auxiliary 
fields in contrast with the 2-form SUGRA 
and the SUSY transformation parameters are not 
constrained at all. We derive the canonical formulation 
of the $N = 2$ theory in such a manner as the relation 
with the usual SUGRA be explicit at least 
in classical level, and show that the algebra of 
the Gauss, $U(1)$ gauge and right-handed SUSY constraints 
form the graded algebra, $G^2SU(2)$. 
Furthermore, we introduce the graded variables associated 
with the $G^2SU(2)$ algebra 
and we rewrite the canonical constraints in a simple form 
in terms of these variables. We quantize the theory 
in the graded-connection representation 
and discuss the solutions of quantum constraints.}} 
\end{center}
\vfill

\newpage

\renewcommand{\thefootnote}{\arabic{footnote}}
\setcounter{section}{0}
\setcounter{equation}{0}
\setcounter{footnote}{0}
\newsection{Introduction}

The nonperturbative canonical treatment of supergravity 
(SUGRA) in terms of the Ashtekar variable \cite{AA} 
was firstly discussed about the simplest $N = 1$ theory 
in \cite{Jac}. In this theory the chiral Lagrangian 
was constructed by using the self-dual connection 
which couples to only a right-handed spin-3/2 field, 
and this Lagrangian has two kinds of right- and left-handed 
supersymmetry (SUSY) invariances in the first-order 
formalism. Therefore two types of the SUSY constraints, 
which generate those SUSY transformations, 
appear in the canonical formulation. F\"ul\"op \cite{Fu} 
and Armand-Ugon {\it et al.} \cite{UGOP} showed that 
in $N = 1$ chiral SUGRA the $SU(2)$ algebra generated 
by the Gauss-law constraint is graded by means 
of the right-handed SUSY constraint. 
All the constraints were also rewritten in a simple form 
in \cite{UGOP} towards a loop representation 
of quantum canonical SUGRA, by using graded connection 
and momentum variables associated with the graded algebra 
which is called the $GSU(2)$ algebra \cite{PR}. 
\footnote{In Ref. \cite{LiSm} it is pointed out that 
the algebra of $GSU(2)$ corresponds to the super 
Lie algebra, $Osp(1/2)$.}
Furthermore the spin network states \cite{Pen} 
of SUGRA was recently constructed in \cite{LiSm} 
based on the representation of this graded algebra. 

The extension of the Ashtekar's canonical formulation 
to $N = 2$ extended SUGRA was mainly developed 
in the context of the 2-form gravity \cite{Sano,KS,Ez}. 
The chiral Lagrangian of $N = 2$ SUGRA was constructed 
in \cite{Sano,KS} based on the $N = 1$ 2-form SUGRA 
\cite{CDJ} with auxiliary fields which are needed 
to write the chiral (2-form) Lagrangian: It was proved 
that the SUSY algebra is not closed at the level of 
transformation algebra on auxiliary fields, but actually 
closes at the level of the canonical formulation. 
On the other hand, in \cite{Ez} the canonical formulation 
of the BF theory as a toplogical field theory \cite{BTH} 
was derived for an appropriate graded algebra of $SU(2)$ 
(which henceforth will be referred to as $G^2SU(2)$ 
\footnote{
The algebra of $G^2SU(2)$ corresponds 
to the super Lie algebra, $Osp(2/2)$ \cite{Marcu}.}
following \cite{Ez}), and it was shown that the $G^2SU(2)$ 
BF theory subject to some algebaic constraints 
can be cast into the $N = 2$ 2-form SUGRA. 

In this paper we reconstruct the Ashtekar's canonical 
formulation of $N = 2$ SUGRA starting from the $N = 2$ 
chiral Lagrangian derived by closely following the method 
employed in the usual SUGRA. 
In Sec. 2 we present the globally $O(2)$ invariant 
Lagrangian of $N = 2$ chiral SUGRA slightly 
modified from the Lagrangian in \cite{TS}. 
In order to get the full graded algebra of the Gauss, 
$U(1)$ gauge and right-handed SUSY constraints, 
we extend in Sec. 3 the internal, global $O(2)$ invariance 
to local one by introducing a cosmological constant 
to the chiral Lagrangian. The resultant Lagrangian 
does not contain any auxiliary fields in contrast 
with the 2-form SUGRA and the SUSY transformation 
parameters are not constrained at all. 
In Sec. 4 we derive the canonical formulation 
of the $N = 2$ theory in such a manner as the relation 
with the usual SUGRA be explicit at least 
in classical level, and we show that the algebra of 
the Gauss, $U(1)$ gauge and right-handed SUSY constraints 
form the graded algebra, $G^2SU(2)$. In Sec. 5 we introduce 
the graded variables associated with the $G^2SU(2)$ algebra 
and rewrite the canonical constraints in a simple form 
in terms of these variables. We quantize the theory 
in the graded-connection representation and discuss 
the solutions of quantum constraints in Sec. 6. 
Our conclusions are included in Sec. 7.

\newsection{The globally $O(2)$ invariant Lagrangian}

Firstly we present the chiral Lagrangian of $N = 2$ 
SUGRA constructed in \cite{TS}. The independent variables 
are a tetrad $e^i_{\mu}$, two (Majorana) Rarita-Schwinger 
fields $\psi^{(I)}_{\mu}$, a Maxwell field $A_{\mu}$ 
and a (complex) self-dual connection $A^{(+)}_{ij \mu}$ 
which satisfies 
$(1/2){\epsilon_{ij}} \! ^{kl} A^{(+)}_{kl \mu}$ 
$= i A^{(+)}_{ij \mu}$. 
\footnote{
Greek letters $\mu, \nu, \dots$, are spacetime indices, 
Lattin letters $i, j, \dots$, are local Lorentz indices 
and $(I), (J), \dots (= (1), (2))$, denote $O(2)$ internal indices. 
We take the Minkowski metric $\eta_{ij} = 
{\rm diag}(-1,+1,+1,+1)$ and the totally antisymmetric tensor 
$\epsilon_{ijkl}$ is normalized as $\epsilon_{0123} = +1$. 
We define $\epsilon_{\mu \nu \rho \sigma}$ 
and $\epsilon^{\mu \nu \rho \sigma}$ as tensor 
densities which take values of $+1$ or $-1$.} 
The $N = 2$ chiral Lagrangian density in terms of these 
variables is written in first-order form as 
%
\ba
{\cal L}^{(+)}_{N = 2} 
\A = \A - {i \over 2} 
        \epsilon^{\mu \nu \rho \sigma} 
        e^i_{\mu} e^j_{\nu} R^{(+)}_{ij \rho \sigma} 
      - \epsilon^{\mu \nu \rho \sigma} 
        \overline \psi^{(I)}_{R \mu} \gamma_\rho 
        D^{(+)}_\sigma \psi^{(I)}_{R \nu} 
      - {e \over 2} (F^{(-)}_{\mu \nu})^2 
\nonu
\A \A + {1 \over{4 \sqrt{2}}} \overline \psi^{(I)}_{\mu} 
        \{ e(F^{\mu \nu} + \hat F^{\mu \nu}) 
        + i \gamma_5 
        (\tilde F^{\mu \nu} + \tilde{\hat F}^{\mu \nu}) \} 
        \psi^{(J)}_{\nu} \epsilon^{(I)(J)} 
\nonu
\A \A + {i \over 8} \epsilon^{\mu \nu \rho \sigma} 
        (\overline \psi^{(I)}_{L \mu} \psi^{(J)}_{R \nu}) 
        \overline \psi^{(K)}_{R \rho} \psi^{(L)}_{L \sigma} 
        \epsilon^{(I)(J)} \epsilon^{(K)(L)}, 
\label{LN2}
\ea
which is globally $O(2)$ invariant. 
Here $e$ denotes ${\rm det}(e^i_{\mu})$, 
$\epsilon^{(I)(J)} = - \epsilon^{(J)(I)}$ 
and $F^{(-) \mu \nu} 
:= (1/2)(F^{\mu \nu} + i e^{-1} \tilde F^{\mu \nu})$ 
with $\tilde F^{\mu \nu} 
= (1/2) \epsilon^{\mu \nu \rho \sigma} F_{\rho \sigma}$. 
The covariant derivative $D^{(+)}_\mu$ and 
the curvature ${R^{(+)ij}}_{\mu \nu}$ are 
\ba
\A \A D^{(+)}_\mu := \partial_\mu + {i \over 2} A^{(+)}_{ij \mu} 
      S^{ij}, \nonu
\A \A {R^{(+)ij}}_{\mu \nu} 
      := 2(\partial_{[\mu} {A^{(+)ij}}_{\nu]} 
      + {A^{(+)i}}_{k [\mu} {A^{(+)kj}}_{\nu]}), 
\ea
while $\hat F_{\mu \nu}$ in the second line of (\ref{LN2}) 
is defined as 
\be
\hat F_{\mu \nu} := F_{\mu \nu} - {1 \over \sqrt{2}} 
\overline \psi^{(I)}_{\mu} \psi^{(J)}_{\nu} 
\epsilon^{(I)(J)}. 
\ee
Note that we have used $(F^{(-)}_{\mu \nu})^2$ 
as the Maxwell kinetic term in Eq. (\ref{LN2}), 
which allows us to rewrite the canonical constraints 
in terms of the graded variables associated with 
the graded algebra, $G^2SU(2)$, as will be 
explained later. In this respect the chiral Lagrangian 
of Eq. (\ref{LN2}) differs from that constructed 
in \cite{TS}. 

The last four-fermion contact term in Eq. (\ref{LN2}) 
is pure imaginary but this term is necessary to reproduce 
the Lagrangian of the usual $N = 2$ SUGRA 
in the second-order formalism. Indeed, if we solve 
the equation 
$\delta {\cal L}^{(+)}_{N = 2}/\delta A^{(+)} = 0$ 
with respect to $A^{(+)}_{ij \mu}$ and use the obtained 
solution in the first two terms in Eq. (\ref{LN2}), 
then those terms give rise to a number of four-fermion 
contact terms, which are complex with the imaginary term 
being written as 
\be
{i \over 8} \epsilon^{\mu \nu \rho \sigma} 
T_{\lambda \mu \nu} T{^{\lambda}}_{\rho \sigma} = 
- {i \over{16}} \epsilon^{\mu \nu \rho \sigma} 
(\overline \psi^{(I)}_{R \mu} \gamma_{\lambda} 
\psi^{(K)}_{R \nu}) 
\overline \psi^{(J)}_{R \rho} \gamma^{\lambda} 
\psi^{(L)}_{R \sigma} \epsilon^{(I)(J)} \epsilon^{(K)(L)}, 
\label{4-Fermi}
\ee
where the torsion tensor is defined by 
$T{^i}_{\mu \nu} = - 2 D_{[\mu} e^i_{\nu]}$ 
with $D_{\mu} e^i_{\nu} = \partial_{\mu} e^i_{\nu} 
+ A{^i}_{j \mu} e^j_{\nu}$. 
The last term in Eq. (\ref{LN2}), on the other hand, 
can be rewritten as 
\ba
\A \A {i \over 8} \epsilon^{\mu \nu \rho \sigma} 
      (\overline \psi^{(I)}_{L \mu} \psi^{(J)}_{R \nu}) 
      \overline \psi^{(K)}_{R \rho} \psi^{(L)}_{L \sigma} 
      \epsilon^{(I)(J)} \epsilon^{(K)(L)} 
\nonu
\A \A = {i \over{16}} \epsilon^{\mu \nu \rho \sigma} 
      (\overline \psi^{(I)}_{R \mu} \gamma_{\lambda} 
      \psi^{(K)}_{R \nu}) 
      \overline \psi^{(J)}_{R \rho} \gamma^{\lambda} 
      \psi^{(L)}_{R \sigma} \epsilon^{(I)(J)} \epsilon^{(K)(L)} 
\ea
by using a Fierz transformation, and exactly cancels with 
the pure imaginary term of Eq. (\ref{4-Fermi}). 
Therefore the ${\cal L}^{(+)}_{N = 2}[{\rm second\ order}]$ 
of $N = 2$ chiral SUGRA is reduced to that of the usual one 
up to imaginary boundary terms; 
namely, we have 
\ba
{\cal L}^{(+)}_{N = 2} [{\rm second\ order}] 
= \A \A {\cal L}_{N = 2 {\rm\ usual\ SUGRA}} 
[{\rm second\ order}] \nonu
\A \A - {1 \over 4} \partial_{\mu} 
\{ \epsilon^{\mu \nu \rho \sigma} 
(\overline \psi^{(I)}_{\nu} \gamma_{\rho} 
\psi^{(I)}_{\sigma} 
+ 2i A_{\nu} \partial_{\rho} A_{\sigma}) \}. 
\label{LN2S}
\ea
Note that a boundary term quadratic in 
the Maxwell field $A_{\mu}$ appears in (\ref{LN2S}) 
since we choose $(F^{(-)}_{\mu \nu})^2$ 
as the kinetic term in Eq. (\ref{LN2}).

\newsection{Gauging the $O(2)$ invariance}

The global $O(2)$ invariance of Eq. (\ref{LN2}) 
can be gauged by introducing a minimal coupling 
for $\psi^{(I)}_{\mu}$ and $A_{\mu}$, 
which automatically requires a spin-3/2 mass-like 
term and a cosmological term in the Lagrangian 
\cite{FrDa}. These three terms are written as 
\ba
{\cal L}_{{\rm cosm}} 
= \A \A {\lambda \over 2} \epsilon^{\mu \nu \rho \sigma} 
\overline \psi^{(I)}_{\mu} \gamma_{\rho} 
\psi^{(J)}_{\nu} A_{\sigma} \epsilon^{(I)(J)} 
\nonu
\A \A - \sqrt{2} i \lambda e \ \overline \psi^{(I)}_{\mu} 
S^{\mu \nu} \psi^{(I)}_{\nu} 
+ 6 \lambda^2 e 
\label{Lcosm}
\ea
with the gauge coupling constant $\lambda$. 
Here the cosmological constant $\Lambda$ is related 
to $\lambda$ as $\Lambda = - 6 \lambda^2$. 
Note that the first term of Eq. (\ref{Lcosm}) is 
comparable with the kinetic term of $\psi^{(I)}_{R \mu}$ 
in Eq. (\ref{LN2}), since this term can be rewritten as 
\be
{\lambda \over 2} \epsilon^{\mu \nu \rho \sigma} 
\overline \psi^{(I)}_{\mu} \gamma_{\rho} 
\psi^{(J)}_{\nu} A_{\sigma} \epsilon^{(I)(J)} 
= \lambda \epsilon^{\mu \nu \rho \sigma} 
\overline \psi^{(I)}_{R \mu} \gamma_{\rho} 
\psi^{(J)}_{R \nu} A_{\sigma} \epsilon^{(I)(J)}. 
\ee
We denote the chiral Lagrangian as the sum of 
Eqs. (\ref{LN2}) and (\ref{Lcosm}); namely, 
\be
{\cal L}^{(+)} 
:= {\cal L}^{(+)}_{N = 2} + {\cal L}_{{\rm cosm}}. 
\label{L+}
\ee
Because of Eq. (\ref{LN2S}), 
the ${\cal L}^{(+)}$ of Eq.(\ref{L+}) in the second-order 
formalism is invariant under the SUSY transformation 
of the usual {\it gauged} $N = 2$ SUGRA \cite{FrDa} 
given by 
\ba
\delta e^i_{\mu} 
= \A \A i \ \overline \alpha^{(I)} 
  \gamma^i \psi^{(I)}_{\mu}, 
\nonu
\delta A_{\mu} 
= \A \A \sqrt{2} \ \epsilon^{(I)(J)} 
  \overline \alpha^{(I)} \psi^{(J)}_{\mu}, 
\nonu
\delta \psi^{(I)}_{\mu} 
= \A \A 2 \{ D_{\mu}[A(e, \psi^{(I)})] \alpha^{(I)} 
  - \lambda \epsilon^{(I)(J)} A_{\mu} \alpha^{(J)} \} 
\nonu
\A \A + {i \over \sqrt{2}} \ \epsilon^{(I)(J)} 
  \left( \hat F_{\mu \nu} \gamma^{\nu} 
  + {i \over 2} e \ \epsilon_{\mu \nu \rho \sigma} 
  \hat F^{\rho \sigma} \gamma^{\nu} \gamma_5 
  \right) \alpha^{(J)} 
\nonu
\A \A - \sqrt{2} i \ \lambda \gamma_{\mu} \alpha^{(I)} 
\label{SUSY-N2}
\ea
with $A_{ij \mu}(e, \psi^{(I)})$ 
in $\delta \psi^{(I)}_{\mu}$ being defined as the sum 
of the Ricci rotation coefficients $A_{ij \mu}(e)$ 
and $K_{ij \mu}$ which is expressed as 
\be
K_{ij \mu} = {i \over 4} 
             (\overline \psi^{(I)}_{[i} 
             \gamma_{\mid \mu \mid} \psi^{(I)}_{j]} 
             + \overline \psi^{(I)}_{[i} 
             \gamma_{\mid j \mid} \psi^{(I)}_{\mu]} 
             - \overline \psi^{(I)}_{[j} 
             \gamma_{\mid i \mid} \psi^{(I)}_{\mu]}). 
\ee
On the other hand, the first-order (i.e.,``off-shell'') 
SUSY invariance of ${\cal L}^{(+)}$ may be 
realized by introducing the right- and left-handed 
SUSY transformations as in the case of 
$N = 1$ chiral SUGRA \cite{Jac,TS2}.

\newsection{The canonical formulation 
of $N = 2$ chiral SUGRA}

Starting with the chiral Lagrangian ${\cal L}^{(+)}$ 
of Eq. (\ref{L+}), let us derive the canonical formulation 
of $N = 2$ chiral SUGRA by means of the (3+1) 
decomposition of spacetime. 
For this purpose we assume that the topology of spacetime 
$M$ is $\Sigma \times R$ for some three-manifold $\Sigma$ 
so that a time coordinate function $t$ is defined on $M$. 
Then the time component of the tetrad can be defined as 
\footnote{Latin letters $a, b, \cdots$ are 
the spatial part of the spacetime indices 
$\mu, \nu, \cdots$, and capital letters 
{\it I, J,} $\cdots$ denote the spatial part 
of the local Lorentz indices {\it i, j,} $\cdots$.}
\be
e^i_t = N n^i + N^a e^i_a. 
\ee
Here $n^i$ is the timelike unit vector orthogonal to 
$e_{ia}$, i.e., $n^i e_{ia} = 0$ and $n^i n_i = - 1$, 
while $N$ and $N^a$ denote the lapse function 
and the shift vector, respectively. 
Furthermore, we give a restriction on the tetrad with 
the choice $n_i = (- 1, 0, 0, 0)$ in order to simplify 
the Legendre transform of Eq. (\ref{L+}). 
Once this choice is made, $e_{Ia}$ becomes tangent to 
the constant $t$ surfaces $\Sigma$ and $e_{0a} = 0$. 
Therefore we change the notation $e_{Ia}$ 
to $E_{Ia}$ below. We also take the spatial restriction 
of the totally antisymmetric tensor 
$\epsilon^{\mu \nu \rho \sigma}$ 
as $\epsilon^{abc} := \epsilon{_t}^{abc}$, 
while $\epsilon^{IJK} := \epsilon{_0}^{IJK}$. 

Under the above gauge condition of the tetrad, 
the (3+1) decomposition of Eq. (\ref{L+}) yields 
the kinetic terms, 
\footnote{
For convenience of calculation we use the two-component 
spinor notation in the canonical formulation. 
Two-component spinor indices $A, B, \dots$, 
and $A', B', \dots$, are raised and lowered 
with antisymmetric spinors $\epsilon^{AB}, \epsilon_{AB}$, 
and their conjugates $\epsilon^{A'B'}, \epsilon_{A'B'}$, 
such that $\psi^A = \epsilon^{AB} \psi_B$ and 
$\varphi_B = \varphi^A \epsilon_{AB}$. 
The Infeld-van der Waerden symbol $\sigma_{iAA'}$ are 
taken in this paper to be $(\sigma_0, \sigma_I) 
:= (-i/\sqrt{2}) (I, \tau_I)$ with $\tau_I$ being 
the Pauli matrices. We also define the symbol 
$\sigma{_{IA}}^B$ (which is called the $SU(2)$ soldering 
form in \cite{AA}) by using $n^{AA'} = n^i \sigma{_i}^{AA'}$ 
as $\sigma{_{IA}}^B := - \sqrt{2} i \sigma_{IAA'} n^{BA'} 
= (i/\sqrt{2})(\tau_I)_{AB}$.}
\be
{\cal L}^{(+)}_{{\rm kin}} 
= \tilde E_I^a \dot {\cal A}{^I}_a 
- \tilde \pi{^{(I)}}{_A}^a \dot \psi{^{(I)A}}_a 
+ {}^+ \tilde \pi^a \dot A_a, 
\label{Lkin}
\ee
where ${\cal A}{^I}_a := - 2 A{^{(+)}}{_0}{^I}_a$ 
and ($\tilde \pi{^{(I)}}{_A}^a$, 
${}^+ \tilde \pi^a$) are defined by 
\footnote{
The derivative for fermionic variables is treated 
as the left derivative unless stated otherwise.}
\ba
\tilde \pi{^{(I)}}{_A}^a := \A \A {{\delta {\cal L}^{(+)}} 
      \over {\delta \dot \psi{^{(I)A}}_a}} 
      = - \sqrt{2} i \ \epsilon^{abc} E^I_c \ 
      \overline \psi{^{(I)A'}}_b \sigma_{IAA'}, 
      \\ 
{}^+ \tilde \pi^a := \A \A {{\delta {\cal L}^{(+)}} 
      \over {\delta \dot A_a}} 
      = \tilde \pi^a + i \ \tilde B^a 
\ea
with 
\ba
\tilde \pi^a := 
      \A \A {e \over {2 N^2}} q^{ab} 
      \{ \ 2 \ (F_{tb} - N^d F_{db}) 
      \nonu
      \A \A - \sqrt{2} \ (\overline \psi^{(I)}_t \psi^{(J)}_b 
      - N^d \overline \psi^{(I)}_d \psi^{(J)}_b) 
      \epsilon^{(I)(J)} \} 
      \nonu
      \A \A - {i \over {2 \sqrt{2}}} \ \epsilon^{abc} 
      \overline \psi^{(I)}_b \gamma_5 \psi^{(J)}_c 
      \epsilon^{(I)(J)}, 
      \label{pi} \\
\tilde B^a = \A \A {1 \over 2} \ \epsilon^{abc} F_{bc}. 
\ea
In Eq. (\ref{pi}) the Majorana spinors $\psi^{(I)}_{\mu}$ 
are used for simplicity. 
On the other hand, the constraints are obtained 
from the variation of ${\cal L}^{(+)}$ with respect to 
Lagrange multipliers. Here we raise, in particular, 
the Gauss, $U(1)$ gauge, right-handed SUSY and left-handed 
SUSY constraints expressed by the canonical variables 
as follows; namely, 
\footnote{
We note that the ${}^+ \tilde \pi^a$ appears 
in Eqs. (\ref{U1}), (\ref{RSUSY}) and (\ref{LSUSY}). 
If we use $(F_{\mu \nu})^2$ as the Maxwell kinetic 
term in Eq. (\ref{LN2}), the $\tilde \pi^a$ 
(and not ${}^+ \tilde \pi^a$) will appear 
in Eq. (\ref{U1}), and it is not possible to rewrite 
the canonical constraints in terms of 
the graded variables.}
\ba
{\cal G}_I := \A \A 
      {{\delta {\cal L}^{(+)}} \over {\delta \Lambda^I_t}} 
      = {\cal D}_a \tilde E_I^a 
      - {i \over \sqrt{2}} \ \tilde \pi{^{(I)}}{_A}^a 
      \sigma{_I}{^A}_B \psi{^{(I)B}}_a = 0, 
      \label{Gauss} \\
g := \A \A 
      {{\delta {\cal L}^{(+)}} \over {\delta A_t}} 
      = \partial_a {}^+ \tilde \pi^a 
      + \lambda \ \psi{^{(I)A}}_a 
      \tilde \pi{^{(J)}}{_A}^a \epsilon^{(I)(J)} = 0, 
      \label{U1} \\
{}^R {\cal S}^{(I)}_A := \A \A 
      {{\delta {\cal L}^{(+)}} \over {\delta \psi{^{(I)A}}_t}} 
      = {\cal D}_a 
      \tilde \pi{^{(I)}}{_A}^a + {1 \over \sqrt{2}} 
      \ {}^+ \tilde \pi^a \ \psi{^{(J)B}}_a 
      \epsilon_{AB} \epsilon^{(I)(J)} 
      \nonu
      \A \A \hspace{18mm} + \lambda \ (2i \tilde E_I^a \sigma{^I}_{AB} 
      \psi{^{(I)B}}_a - \tilde \pi{^{(J)}}{_A}^a 
      A_a \epsilon^{(I)(J)}) = 0, 
      \label{RSUSY} \\
{}^L {\cal S}^{(I)}_A := \A \A 
      {{\delta {\cal L}^{(+)}} \over {\delta \rho{^{(I)A}}_t}} 
      \nonu
      = \A \A - \sqrt{2} 
      \ \tilde E_I^a \tilde E_J^b {(\sigma^I \sigma^J)}{_A}^B 
      \Bigg[ \ 2({\cal D}_{[a} \psi{^{(I)C}}_{b]} 
      - \lambda \ A_{[a} \psi{^{(J)C}}_{b]} 
      \epsilon^{(I)(J)}) \epsilon_{BC} 
      \nonu
      \A \A \left. + {i \over \sqrt{2}} 
      \lambda \ \epsilon_{abc} \tilde \pi{^{(I)}}{_B}^c \right] 
      \nonu
      \A \A + {i \over 2} E^{-2} \epsilon_{def} \epsilon_{agh} 
      {(\sigma^I \sigma^J \sigma^K \sigma^L)}{_A}^B 
      \epsilon^{(I)(J)} \tilde E_I^e \tilde E_J^f 
      \tilde E_K^g \tilde E_L^h 
      \tilde \pi{^{(J)}}{_B}^d 
      \nonu
      \A \A \times \left[ \epsilon^{abc} \left\{ F_{bc} 
      + {1 \over \sqrt{2}} 
      \ \epsilon_{CD} (\psi{^{(K)C}}_b \psi{^{(L)D}}_c) 
      \epsilon^{(K)(L)} \right\} 
      + i \ {}^+ \tilde \pi^a \right] = 0, 
\label{LSUSY}
\ea
where the Lagrange multipliers, $\Lambda^I_t$ 
and $\rho{^{(I)A}}_t$, are defined by 
\be
\Lambda^I_t := - 2 A^{(+)}{_0}{^I}_t, \ \ \ \ \ 
\rho{^{(I)A}}_t := E^{-1} \overline \psi^{(I)}_{A't} \ 
               n^{AA'}, 
\ee
and the covariant derivatives on $\Sigma$ are 
\ba
\A \A {\cal D}_a \tilde E_I^a 
:= \partial_a \tilde E_I^a + i \epsilon_{IJK} 
{\cal A}{^J}_a \tilde E^{Ka}, 
\nonu
\A \A {\cal D}_a \tilde \pi{^{(I)}}{_A}^a 
:= \partial_a \tilde \pi{^{(I)}}{_A}^a 
- {i \over \sqrt{2}} {\cal A}{_A}{^B}_a 
\tilde \pi{^{(I)}}{_B}^a. 
\ea
The left-handed SUSY constraint of Eq. (\ref{LSUSY}) 
is not polynomial because of the factor 
$E^{-2}$, but the rescaled $E^2 ({}^L {\cal S}^{(I)}_A)$ 
becomes polynomial because 
\be
E^2 = {1 \over 6} \epsilon_{abc} \epsilon^{IJK} 
      \tilde E_I^a \tilde E_J^b \tilde E_K^c. 
\ee
The above canonical constraints (except for the Gauss 
constraint) have expressions different from 
those for the $N = 2$ 2-form SUGRA \cite{Sano,KS}. 
This seems to originate from difference in the definition 
of momentum variables, in particular the momentum conjugate 
to the Maxwell field. 

Now, by using the non-vanishing Poisson brackets 
\footnote{The Poisson brackets are defined 
for canonical variables $(q^i, \tilde p_i)$ 
by using the right and left derivatives as 
$\{ F, G \} := \int d^3 z [(\delta^R F/\delta q^i(z)) 
(\delta^L G/\delta \tilde p_i(z)) 
- (-1)^{\mid i \mid} (\delta^R F/\delta \tilde p_i(z)) 
(\delta^L G/\delta q^i(z))]$ 
with $\mid i \mid = 0$ for an even (commuting) $q^i$ 
while $\mid i \mid = 1$ for an odd (anticommuting) $q^i$.}
among the canonical variables, 
\ba
\A \A \{ {\cal A}{^I}_a(x), \tilde E{_J}^b(y) \} 
= \delta^I_J \delta_a^b \delta^3(x - y), 
\nonu
\A \A \{ \psi{^{(I)A}}_a(x), 
\tilde \pi{^{(J)}}{_B}^b(y) \} 
= - \delta^{(I)(J)} \delta_B^A \delta_a^b \delta^3(x - y), 
\nonu
\A \A \{ A_a(x), {}^+ \tilde \pi^b(y) \} 
= \delta_a^b \delta^3(x - y), 
\ea
we show that the Gauss, $U(1)$ gauge and right-handed 
SUSY constraints of Eqs. (\ref{Gauss})-(\ref{RSUSY}) 
form the graded algbra, $G^2SU(2)$. 
In fact, if we define the smeared functions, 
\ba
\A \A {\cal G}_I [\Lambda^I] := \int_{\Sigma} d^3 x \ 
                             \Lambda^I {\cal G}_I, 
                             \nonu
\A \A g [a] := \int_{\Sigma} d^3 x \ 
               a \ g, 
               \nonu
\A \A {}^R {\cal S}^{(I)}_A [\xi^{(I)A}] 
               := \int_{\Sigma} d^3 x \ 
               \xi^{(I)A} \ {}^R {\cal S}^{(I)}_A 
\ea
for convenience of the calculation, 
the Poisson brackets of 
${\cal G}_I, g$ and ${}^R {\cal S}^{(I)}_A$ 
are obtained as 
\ba
\A \A \{ {\cal G}_I [\Lambda^I], \ {\cal G}_J [\Gamma^J] \} 
      = {\cal G}_I [\Lambda'^I], 
\nonu
\A \A \{ {\cal G}_I [\Lambda^I], \ g [a] \} 
      = 0 = \{ g [a], \ g [b] \}, 
\nonu
\A \A \{ {\cal G}_I [\Lambda^I], 
      \ {}^R {\cal S}^{(I)}_A [\xi^{(I)A}] \} 
      = {}^R {\cal S}^{(I)}_A [\xi'^{(I)A}], 
\nonu
\A \A \{ g [a], \ {}^R {\cal S}^{(I)}_A [\xi^{(I)A}] \} 
      = \lambda \ {}^R {\cal S}^{(I)}_A [\xi''^{(I)A}], 
\nonu
\A \A \{ {}^R {\cal S}^{(I)}_A [\xi^{(I)A}], 
      \ {}^R {\cal S}^{(J)}_B [\eta^{(J)B}] \} 
      = \lambda \ {\cal G}_I [\Lambda''^I] + g [a'] 
\label{gLie}
\ea
with the parameters, $\Lambda'^I, \Lambda''^I, 
\xi'^{(I)A}, \xi''^{(I)A}$ and $a'$, being defined as 
\ba
\A \A \Lambda'^I := i \ \epsilon^{IJK} 
      \Lambda_J \Gamma_K, 
\nonu
\A \A \Lambda''^I := 2i \ \xi^{(I)A} \eta^{(J)B}
      \sigma{^I}_{AB} \delta^{(I)(J)}, 
\nonu
\A \A \xi'^{(I)A} := {i \over \sqrt{2}} \ 
      \Lambda^I \xi^{(I)B} \sigma{_{IB}}^A, 
\nonu
\A \A \xi''^{(I)A} := - a \ \xi^{(J)A} 
      \epsilon^{(I)(J)}, 
\nonu
\A \A a' := {1 \over \sqrt{2}} \ \xi^{(I)A} \eta^{(J)B}
      \epsilon_{AB} \epsilon^{(I)(J)}. 
\ea
The algebra of Eq. (\ref{gLie}) coincides with 
the graded algebra, $G^2SU(2)$, which was first introduced 
in \cite{Ez} in the framework of the BF theory.

\newsection{The graded variables associated 
with the $G^2SU(2)$ algebra}

In $N = 1$ chiral SUGRA, the graded variables of $GSU(2)$ 
was introduced based on the graded algebra which is 
satisfied by the Gauss and right-handed SUSY constraints 
\cite{Fu,UGOP}. These graded variables simplify 
the expressions of all the canonical constraints, 
and therefore it becomes easier to find exact solutions 
of quantum constraints \cite{UGOP}. 

In order to introduce the graded variables 
of $G^2SU(2)$ in $N = 2$ chiral SUGRA, 
let us define the generators 
\be
J_{\hat i} := (J_I, J_{\alpha}, J_8), 
\ee
which satisfy the same algebra as that of the constraints 
\be
C_{\hat i} := ({\cal G}_I, 
              {}^R {\cal S}_{\alpha}, g_8), 
\ee
where $({}^R {\cal S}_{\alpha}, g_8)$ stand for 
$({}^R {\cal S}^{(I)}_A, g)$, and the index $\hat i$ 
runs over $(I, \alpha, 8)$ with $\alpha := (I)A$. 
Namely, we suppose that the $J_{\hat i}$ satisfy 
the $G^2SU(2)$ algebra 
\be
[J_{\hat i}, J_{\hat j} \} 
= f{_{\hat i \hat j}}^{\hat k} J_{\hat k} 
\label{Jalg}
\ee
with $f{_{\hat i \hat j}}^{\hat k}$ being the structure 
constant determined from Eq. (\ref{gLie}). 
The fundamental representation of this algebra 
is given by 
\ba
\A \A J_1 = {1 \over 2} 
            \pmatrix{
            0 & 1 & 0 & 0 \cr
            1 & 0 & 0 & 0 \cr
            0 & 0 & 0 & 0 \cr
            0 & 0 & 0 & 0 \cr
            }, \ \ 
      J_2 = {1 \over 2} 
            \pmatrix{
            0 & -i & 0 & 0 \cr
            i &  0 & 0 & 0 \cr
            0 &  0 & 0 & 0 \cr
            0 &  0 & 0 & 0 \cr
            }, \ \ 
      J_3 = {1 \over 2} 
            \pmatrix{
            1 &  0 & 0 & 0 \cr
            0 & -1 & 0 & 0 \cr
            0 &  0 & 0 & 0 \cr
            0 &  0 & 0 & 0 \cr
            }, 
            \nonu
\A \A J^{(1)}_1 = \sqrt{{\lambda \over \sqrt{2}}} \ i 
            \pmatrix{
            0 & 0 &  0 & 0 \cr
            0 & 0 & -1 & 0 \cr
            1 & 0 &  0 & 0 \cr
            0 & 0 &  0 & 0 \cr
            }, \ \ \ 
      J^{(1)}_2 = \sqrt{{\lambda \over \sqrt{2}}} \ i 
            \pmatrix{
            0 & 0 & 1 & 0 \cr
            0 & 0 & 0 & 0 \cr
            0 & 1 & 0 & 0 \cr
            0 & 0 & 0 & 0 \cr
            }, 
            \nonu
\A \A J^{(2)}_1 = \sqrt{{\lambda \over \sqrt{2}}} \ i 
            \pmatrix{
            0 & 0 & 0 &  0 \cr
            0 & 0 & 0 & -1 \cr
            0 & 0 & 0 &  0 \cr
            1 & 0 & 0 &  0 \cr
            }, \ \ \ 
      J^{(2)}_2 = \sqrt{{\lambda \over \sqrt{2}}} \ i 
            \pmatrix{
            0 & 0 & 0 & 1 \cr
            0 & 0 & 0 & 0 \cr
            0 & 0 & 0 & 0 \cr
            0 & 1 & 0 & 0 \cr
            }, 
            \nonu
\A \A J_8 = \lambda 
            \pmatrix{
            0 & 0 & 0 &  0 \cr
            0 & 0 & 0 &  0 \cr
            0 & 0 & 0 & -1 \cr
            0 & 0 & 1 &  0 \cr
            }. 
\label{frep}
\ea
The supertrace \cite{DeW} of the bilinear forms, 
${\rm STr}(J_{\hat i} J_{\hat j})$, is given by 
\ba
\A \A {\rm STr}(J_I J_J) 
      = {1 \over 2} \ \delta_{IJ}, \ \ 
      {\rm STr}(J_{\alpha} J_{\beta}) 
      = {\rm STr}(J^{(I)}_A J^{(J)}_B) 
      = \sqrt{2} \lambda \epsilon_{AB} 
      \delta^{(I)(J)}, 
      \nonu
\A \A {\rm STr}(J_8 J_8) = 2 \lambda^2, \ \ 
      {\rm STr}(J_I J_{\alpha}) 
      = 0 = {\rm STr}(J_I J_8) 
      = {\rm STr}(J_{\alpha} J_8), 
\label{STr}
\ea
where the supertrace for a $G^2SU(2)$ matrix $M$ 
is defined by ${\rm STr}(M) = M_{11} + M_{22} 
- M_{33} - M_{44}$. We introduce the metric 
$g_{\hat i \hat j}$ for $G^2SU(2)$, 
which is of block-diagonal form, by 
\be
g_{\hat i \hat j} := 2 \ {\rm STr}(J_{\hat i} J_{\hat j}) 
                  = (\delta_{IJ}, \ 2 \sqrt{2} \lambda 
                    \epsilon_{AB} \delta^{(I)(J)}, 
                    \ 4 \lambda^2). 
\ee
Here we nomalize $g_{\hat i \hat j}$ so that 
the condition $g_{IJ} = \delta_{IJ}$ be satisfied. 
The inverse $g^{\hat i \hat j}$ is given by 
\be
g^{\hat i \hat j} = \left( \delta^{IJ}, 
                    - {1 \over {2 \sqrt{2} \lambda}} 
                    \epsilon^{AB} \delta^{(I)(J)}, 
                    {1 \over {4 \lambda^2}} \right). 
\ee
We shall lower or raise the index $\hat i$ by using 
$g_{\hat i \hat j}$ and $g^{\hat i \hat j}$. 
For example, the $J^{\hat i}$ with upper index $\hat i$ 
is defined by 
$J^{\hat i} := g^{\hat i \hat j} J_{\hat j}$. 

As is seen from Eq. (\ref{Lkin}), the sets of the fields, 
\ba
\A \A {\bf{\cal A}}{^{\hat i}}_a 
:= ({\cal A}{^I}_a, \ \psi{^{\alpha}}_a := \psi{^{(I)A}}_a, 
\ A^8_a := A_a), 
\label{gvA} \\
\A \A \tilde {\bf{\cal E}}{_{\hat i}}^a = (\tilde E{_I}^a, 
\ - \tilde \pi{_{\alpha}}^a := - \tilde \pi{^{(I)}}{_A}^a, 
\ {}^+ \tilde \pi_8^a := {}^+ \tilde \pi^a), 
\label{gvE}
\ea
play the role of the coordinate variables and 
their conjugate momenta, respectively. 
We shall refer to ${\bf{\cal A}}{^{\hat i}}_a$ and 
$\tilde {\bf{\cal E}}{_{\hat i}}^a$ as 
the graded connection and the graded momentum, 
respectively. 

Now let us define the graded variables ${\bf{\cal A}}_a$ 
and $\tilde {\bf{\cal E}}^a$: 
\be
{\bf{\cal A}}_a := 
{\bf{\cal A}}{^{\hat i}}_a J_{\hat i}, \ \ \ \ 
\tilde {\bf{\cal E}}^a 
:= \tilde {\bf{\cal E}}{_{\hat i}}^a J^{\hat i}. 
\ee
Then the kinetic terms of Eq. (\ref{Lkin}) are 
expressed as a simple form, 
$2 \ {\rm STr} (\tilde {\bf{\cal E}}^a \dot 
{\bf{\cal A}}_a)$, due to the relation 
${\rm STr}(J^{\hat i} J_{\hat j}) 
= (1/2) \delta^{\hat i}_{\hat j}$. 
Furthermore, we can show that 
the divergence of $\tilde {\bf{\cal E}}^a$ 
can be rewritten as 
\ba
{\bf{\cal D}}_a \tilde {\bf{\cal E}}^a 
:= \A \A \partial_a \tilde {\bf{\cal E}}^a 
   + [ {\bf{\cal A}}_a, \tilde {\bf{\cal E}}^a \} 
\nonu
= \A \A \partial_a 
  \tilde {\bf{\cal E}}{_{\hat i}}^a J^{\hat i} 
  + {\bf{\cal A}}{^{\hat i}}_a 
  \tilde {\bf{\cal E}}{_{\hat j}}^a 
  [ J_{\hat i}, J^{\hat j} \} 
\nonu
= \A \A C_{\hat i} J^{\hat i}. 
\ea
Therefore, the Gauss, $U(1)$ gauge 
and right-handed SUSY constraints 
are unified into the $G^2SU(2)$ Gauss law 
\be
{\bf{\cal D}}_a \tilde {\bf{\cal E}}^a = 0. 
\label{GGauss}
\ee

The left-handed SUSY constraint can also be expressed 
by using the graded variables, 
${\bf{\cal A}}{^{\hat i}}_a$ and 
$\tilde {\bf{\cal E}}{_{\hat i}}^a$ as in the case of 
$N = 1$ chiral SUGRA \cite{UGOP}. 
Indeed, ${}^L {\cal S}^{(I)}_A$ of Eq. (\ref{LSUSY}) 
can be rewritten as 
\ba
{}^L {\cal S}^{(I)}_A 
= \A \A \lambda^{-1} 
  (C^{(I)}{_{\! \! \! \! A}}^{\ \hat i \hat j \hat k} 
  \ \epsilon_{abc} \ 
  \tilde {\bf{\cal E}}{_{\hat i}}^a 
  \tilde {\bf{\cal E}}{_{\hat j}}^b 
  \nonu
  \A \A + \lambda^{-1} E^{-2} 
  C^{(I)}{_{\! \! \! \! A}}^{\ \hat l \hat m \hat n 
  \hat i \hat j \hat k} 
  \ \epsilon_{def} \epsilon_{cgh} \ 
  \tilde {\bf{\cal E}}{_{\hat l}}^e 
  \tilde {\bf{\cal E}}{_{\hat m}}^f 
  \tilde {\bf{\cal E}}{_{\hat n}}^g 
  \tilde {\bf{\cal E}}{_{\hat i}}^h 
  \tilde {\bf{\cal E}}{_{\hat j}}^d ) 
  \ ({\bf{\cal B}}{_{\hat k}}^c \pm 2i \lambda^2 
  \tilde {\bf{\cal E}}{_{\hat k}}^c) 
  \nonu
= \A \A 0, 
\label{G-LSUSY}
\ea
where $C^{(I)}{_{\! \! \! \! A}}^{\ \hat i \hat j \hat k}$ 
and $C^{(I)}{_{\! \! \! \! A}}^{\ \hat l \hat m \hat n 
\hat i \hat j \hat k}$ are defined by 
\ba
\A \A C^{(I)}{_{\! \! \! \! A}}^{\ IJ \alpha} 
      := - {1 \over 2} (\sigma^I \sigma^J){_A}^B 
      \delta^{(I)(J)}, \\
\A \A C^{(I)}{_{\! \! \! \! A}}^{\ LMNI \alpha 8}
      := - {i \over 4} 
      (\sigma^L \sigma^M \sigma^N \sigma^I){_A}^B 
      \epsilon^{(I)(J)} 
\ea
while all other components vanish, 
and the signatures $\pm$ in the parenthesis 
of Eq. (\ref{G-LSUSY}) are taken to be $+$ for 
$\hat k = I, 8$ while $-$ for $\hat k = \alpha$. 
In Eq. (\ref{G-LSUSY}) we have also defined 
${\bf{\cal B}}^{\hat i a}$ as 
${\bf{\cal B}}^{\hat i a} = (1/2) \epsilon^{abc} 
{\bf{\cal F}}{^{\hat i}}_{bc}$ with 
\ba
{\bf{\cal F}}{^{\hat i}}_{ab} 
:= \A \A 2 \ \partial_{[a} {\bf{\cal A}}{^{\hat i}}_{b]} 
   + f{_{\hat j \hat k}}^{\hat i} 
   {\bf{\cal A}}{^{\hat j}}_a 
   {\bf{\cal A}}{^{\hat k}}_b 
   \nonu
=  \A \A \Bigg( {\cal F}{^I}_{ab} + 2i \lambda 
   \psi{^{(I)A}}_{[a} \sigma{^I}_{\mid AB \mid} 
   \psi{^{(I)B}}_{b]}, 
   \nonu
   \A \A 2( {\cal D}_{[a} \psi{^{(J)A}}_{b]} 
   - \lambda A_{[a} \psi{^{(J)A}}_{b]}) 
   \epsilon^{(I)(J)}, 
   \nonu
   \A \A \left. F_{ab} + {1 \over \sqrt{2}} \epsilon_{AB} 
   (\psi{^{(I)A}}_{[a} \psi{^{(J)B}}_{b]}) 
   \epsilon^{(I)(J)} \right), 
\ea
where ${\cal F}{^I}_{ab} := 2 \ \partial_{[a} 
{\cal A}{^I}_{b]} + i \ \epsilon{^I}_{JK} 
{\cal A}{^J}_a {\cal A}{^K}_b$.

\newsection{Quantization in the graded-connection 
representation}

In this section, we consider the canonical quantization 
of $N = 2$ chiral SUGRA in the graded-connection 
representation: Namely, quantum states are represented 
by wavefunctionals $\Psi[{\bf{\cal A}}]$, 
and operators of the graded variables 
$(\hat {\bf{\cal A}}{^{\hat i}}_a, 
\hat{\tilde {\bf{\cal E}}}{_{\hat i}}^a)$ 
act on $\Psi[{\bf{\cal A}}]$ as 
\be
\hat {\bf{\cal A}}{^{\hat i}}_a \Psi[{\bf{\cal A}}] 
= {\bf{\cal A}}{^{\hat i}}_a \Psi[{\bf{\cal A}}], 
\ \ \ \ \hat{\tilde {\bf{\cal E}}}{_{\hat i}}^a 
\Psi[{\bf{\cal A}}] 
= \mp i {\delta \over {\delta {\bf{\cal A}}{^{\hat i}}_a}} 
\Psi[{\bf{\cal A}}], 
\ee
where the signatures in front of the derivative 
$(\delta/{\delta {\bf{\cal A}}{^{\hat i}}_a})$ 
are taken to be $-$ for $\hat i = I, 8$ while 
$+$ for $\hat i = \alpha$. 
In $N = 1$ chiral SUGRA \cite{UGOP}, 
there are two main results about solutions of quantum 
constraints in the graded-connection representation 
of $GSU(2)$ as in the case of pure gravity 
\cite{JaSm,KoBGP}. 
One is that with the factor-ordering of the triads 
to the right Wilson loops of the graded connection 
are annihilated by the quantum Gauss and right-handed 
SUSY constraints, and also annihilated by 
the quantum left-handed SUSY contraint for smooth loops 
(which do not have kinks or intersections). 
The other is that with ordering the triads to the left 
in the left-handed SUSY constraint the exponential 
of the Chern-Simons form built with the graded connection 
solves all quantum constraints with a cosmological 
constant. 

Let us examine quantum constraints for the above two cases 
in $N = 2$ chiral SUGRA, for which the left-handed SUSY 
constraint of Eq. (\ref{G-LSUSY}) involves a nonpolynomial 
factor $E^{-2}$ like the Hamiltonian constraint 
in the Einstein-Maxwell theory in the Ashtekar variable 
\cite{GaPu}. Firstly, we consider Wilson loops 
of the graded connection for $G^2SU(2)$, 
\be
W_{\gamma}[{\bf{\cal A}}] 
= {\rm STr} \left[ P \ {\rm exp} 
\left( \oint_{\gamma} d y^a {\bf{\cal A}}_a (y) 
\right) \right], 
\label{Wloop}
\ee
where the path ordered exponential is used 
and $\gamma$ denotes loops on $\Sigma$. 
The Wilson loops of Eq. (\ref{Wloop}) are $G^2SU(2)$ 
invariant so that they are annihilated by 
the quantum $G^2SU(2)$ Gauss law (\ref{GGauss}), 
and for smooth loops they are also annihilated 
by the rescaled left-handed SUSY constraint, 
$\hat E^2 ({}^L \hat {\cal S}^{(I)}_A)$. 
However, if the rescaled constraint 
is to be equivalent with the original constraint, 
then Eq. (\ref{Wloop}) fails to be solutions 
in $N = 2$ chiral SUGRA as stated in \cite{GaPu} 
since the operator $\hat E^2$ annihilates 
$W_{\gamma}[{\bf{\cal A}}]$. 

Secondly, we consider the exponential of the Chern-Simons 
form built with the graded connection of $G^2SU(2)$, 
\be
\Psi_{{\rm CS}}[{\bf{\cal A}}] = {\rm exp} \left[ 
- {1 \over {2 \lambda^2}} \int_{\Sigma} d^3 x 
\epsilon^{abc} \ {\rm STr} \left( 
{\bf{\cal A}}_a \partial_b {\bf{\cal A}}_c 
+ {2 \over 3} {\bf{\cal A}}_a {\bf{\cal A}}_b 
{\bf{\cal A}}_c \right) \right]. 
\label{N2CS}
\ee
This is an exact state functional that solves all quantum 
constraints of $N = 2$ chiral SUGRA: 
In fact, Eq. (\ref{N2CS}) is annihilated by the $G^2SU(2)$ 
Gauss law because of a $G^2SU(2)$ invariance 
for $\Psi_{{\rm CS}}[{\bf{\cal A}}]$, 
while it is annihilated by the quauntum 
left-handed SUSY constraint for Eq. (\ref{G-LSUSY}) 
with the factor-ordering of the triads to the left; 
namely, 
\be
{}^L \hat {\cal S}^{(I)}_A \Psi_{{\rm CS}}[{\bf{\cal A}}] 
= \dots \times \left( {\bf{\cal B}}{_{\hat k}}^c + 2 \lambda^2 
  {\delta \over {\delta {\bf{\cal A}}{^{\hat k}}_c}} 
  \right) \Psi_{{\rm CS}}[{\bf{\cal A}}] = 0, 
\ee
since $(\delta/\delta {\bf{\cal A}}{^{\hat k}}_c) 
\Psi_{{\rm CS}}[{\bf{\cal A}}] = (-1/2 \lambda^2) 
{\bf{\cal B}}{_{\hat k}}^c \Psi_{{\rm CS}}[{\bf{\cal A}}]$. 
The $\Psi_{{\rm CS}}[{\bf{\cal A}}]$ of Eq. (\ref{N2CS}) 
coincides with the $N = 2$ supersymmetric Chern-Simons 
solution obtained in \cite{Sano,Ez}.

\newsection{Conclusions}

In this paper we have reconstructed the Ashtekar's canonical 
formulation of $N = 2$ SUGRA starting from the $N = 2$ 
chiral Lagrangian derived by closely following the method 
employed in the usual SUGRA. 
We have modified the Maxwell kinetic term as 
$(F^{(-)}_{\mu \nu})^2$ in the globally $O(2)$ invariant 
Lagrangian obtained in \cite{TS}. 
In addition we have gauged $O(2)$ invariance of the Lagrangian 
so that we have obtained the full graded algebra, 
$G^2SU(2)$, of the Gauss, $U(1)$ gauge and right-handed 
SUSY constraints in the canonical formulation. 
The left-handed SUSY constraint has the nonpolynomial 
factor $E^{-2}$ as in the case of the Einstein-Maxwell 
theory in the Ashtekar variable \cite{GaPu}. 

We have introduced the graded variables 
$({\bf{\cal A}}_a, \tilde {\bf{\cal E}}^a)$ 
associated with the $G^2SU(2)$ algebra 
and showed that the $G^2SU(2)$ Gauss law, 
${\bf{\cal D}}_a \tilde {\bf{\cal E}}^a = 0$, 
coincides with the Gauss, $U(1)$ gauge and right-handed 
SUSY constraints. We have also rewritten the left-handed 
SUSY constraint in terms of the graded variables. 
Based on the representation in which the graded 
connection is diagonal, we have examined the solutions 
of quantum constraints obtained in $N = 1$ chiral SUGRA 
\cite{UGOP}; namely, Wilson loops 
of the graded connection, $W_{\gamma}[{\bf{\cal A}}]$, 
and the exponential of the Chern-Simons form 
built with the graded connection, 
$\Psi_{{\rm CS}}[{\bf{\cal A}}]$. 
If the left-handed SUSY constraint rescaled by $E^2$ 
is to be equivalent with the original constraint, 
then the Wilson loops of $W_{\gamma}[{\bf{\cal A}}]$ 
fail to be solutions in $N = 2$ chiral SUGRA 
since the operator $\hat E^2$ annihilates 
$W_{\gamma}[{\bf{\cal A}}]$. 
On the other hand, the exponential of the Chern-Simons form, 
$\Psi_{{\rm CS}}[{\bf{\cal A}}]$, is 
an exact state functional that solves all quantum constraints 
of $N = 2$ chiral SUGRA. 
This solution was first derived in \cite{Sano} 
in the 2-form SUGRA, and later given in \cite{Ez} 
based on the BF theory. 
In this paper we have obtained the same result starting 
from the chiral Lagrangian which is closely related 
to the usual SUGRA. 

The extension to higher $N$ SUGRA is now investigated.

{\large{\bf{Acknowledgments}}} 

I would like to thank 
the members of Physics Department at Saitama University 
for discussions and encouragements. 
This work was supported by the High-Tech Research Center 
of Saitama Institute of Technology.



\end{document}